\documentclass[apl,twocolumn,amsmath,amssymb,superscriptaddress,floatfix]{revtex4-1}

\usepackage{graphicx}
\usepackage{dcolumn}
\usepackage{bm}
\usepackage[sort&compress]{natbib}
\usepackage{xspace}
\usepackage{amsmath}
\usepackage{amssymb}
\usepackage{epsfig}

\newcommand{\mum}{ \,\mu\text{m}}

\newcommand{\lom} {L_{\text{OM}}}

\newcommand{\SiN}{\text{Si}_\text{3}\text{N}_\text{4}}

\begin{document}

\title{$\SiN$ optomechanical crystals in the resolved-sideband regime}

\author{M. Davan\c co} \email{mdavanco@nist.gov}
\affiliation{Center for Nanoscale Science and Technology,
National Institute of Standards and Technology, Gaithersburg, MD
20899, USA}
\affiliation{Department of Applied Physics, California Institute of Technology, Pasadena, CA 91125, USA}
\author{S. Ates}
\altaffiliation[Currently with ]{The National Research Institute of Electronics and Cryptology, The Scientific and Technological Research Council of Turkey, Gebze 41400, Turkey }
\affiliation{Center for Nanoscale Science and Technology,
National Institute of Standards and Technology, Gaithersburg, MD
20899, USA}
\affiliation{Maryland NanoCenter, University of Maryland, College
Park, MD}
\author{Y. Liu}
\altaffiliation[Currently with the ]{Department of Mechanical Engineering, Worcester Polytechnic Institute,
100 Institute Road,
Worcester, MA 01609-2280
}
\affiliation{Center for Nanoscale Science and Technology,
National Institute of Standards and Technology, Gaithersburg, MD
20899, USA}
\affiliation{Maryland NanoCenter, University of Maryland, College
Park, MD}
\author{K. Srinivasan}
\affiliation{Center for Nanoscale Science and Technology, National
Institute of Standards and Technology, Gaithersburg, MD 20899, USA}

\date{\today}

\begin{abstract}
We demonstrate sideband-resolved $\SiN$ optomechanical crystals supporting $10^5$ quality factor optical modes at 980 nm, coupled to $\approx 4$~GHz frequency mechanical modes with quality factors of $\approx 3000$. Optomechanical electromagnetically induced transparency and absorption are observed at room temperature and in atmosphere with intracavity photon numbers of the order of $10^4$.
\end{abstract}

 \maketitle

Silicon optomechanical crystals (OMCs) have been used to demonstrate important steps on the road to performing chip-scale quantum mechanical
operations with individual phonons of a mesoscopic object. In~\cite{ref:safavi-naeini4}, coherent interaction between cavity photons and phonons was used to demonstrate electromagnetically induced transparency (EIT) mediated by a mechanical resonance; radiation-pressure
cooling of one OMC mechanical resonance to its ground state was shown in~\cite{ref:Chan_Painter_ground_state}. Silicon OMCs are characterized by several properties that have enabled these demonstrations.  They support co-localized GHz mechanical and near-infrared optical resonances that interact via radiation pressure, with large zero-point optomechanical coupling rates ($g_0$) between cavity photons and phonons. High quality factor ($Q_o>10^6$) optical resonances can be achieved, with linewidth $\kappa$ smaller than the few-GHz mechanical resonance frequency $f_m$. This places the system in the resolved-sideband regime ($\omega_m/\kappa>1$, $\omega_m=2\pi f_m$), in which efficient radiation-pressure dynamical back-action effects can be observed~\cite{ref:Kippenerbg_Vahala_Science}. Lastly, mechanical resonances with high quality factors ($Q_m>10^5$) are achievable, necessary for pronounced optomechanical interactions, as quantified by the cooperativity parameter $C=4Ng_0^2/\kappa\gamma_i$ ($N$ is the intracavity photon number and $\gamma_i$ is the mechanical linewidth). Large $C$ is necessary for effective laser cooling~\cite{ref:Kippenerbg_Vahala_Science,ref:Chan_Painter_ground_state} and coherent photon-phonon exchange~\cite{ref:safavi-naeini3,ref:Verhagen_Kippenberg_Nature}, and is achieved in Si OMCs due to their aforementioned small $\gamma_i$, small $\kappa$, and large $g_0$. $C$ also scales linearly with the intracavity photon population, and thus large $N$ provides another route to appreciable $C$, as observed in systems such as silica microtoroids~\cite{ref:Verhagen_Kippenberg_Nature} and superconducting cavity electromechanical circuits~\cite{ref:Teufel_strong_coupling}. This strategy has found limited use in Si OMCs because of silicon's large two-photon absorption coefficient, which leads to reduced $Q_o$ for large $N$.

As an alternative to Si, $\SiN$ is a potentially advantageous material for OMCs: it does not exhibit two-photon absorption at relevant wavelengths; like Si, it lends itself to chip-scale nanofabrication with mature techniques; stoichiometric $\SiN$ films under tensile
stress have been shown to support high optical and mechanical $Q$s ($\approx10^6$~\cite{ref:Verbridge,ref:Bauters_Bowers_Si3N_wgs}); and $\SiN$ is transparent over the visible and near-infrared bands, contrasting with Si, which is opaque below 1~$\mu$m. The latter is crucial for interaction with a variety of quantum optical systems that operate at shorter wavelengths, such as semiconductor quantum dots, single molecules in organic crystals, nitrogen-vacancy defects in diamond, and trapped atoms and ions.

\begin{figure}[h]
\centerline{\includegraphics[width=9cm,trim=0 5 0 15]{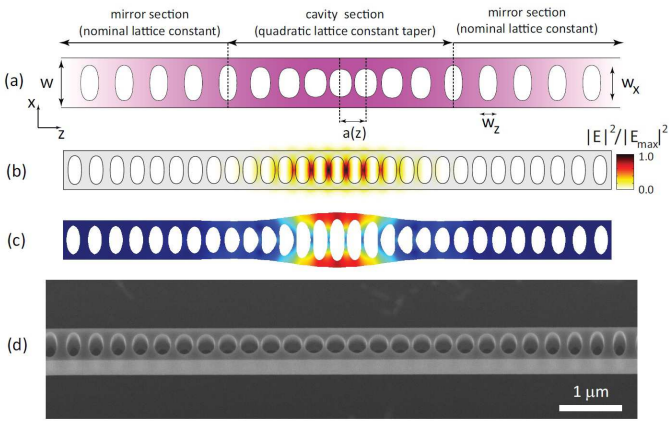}}
\caption{(a) Optomechanical crystal geometry. (b) First-order optical resonance. (c) First-order breathing mechanical mode. (d) SEM of fabricated OMC.}
\label{Fig_1}
\end{figure}

Here, we design and experimentally investigate $\SiN$-based OMCs in the resolved-sideband limit for light in the 980~nm wavelength range. The sideband resolution achieved was sufficient for observing, at room temperature and atmospheric pressure, EIT mediated by a $\approx4$~GHz mechanical mode. Our OMC consists of a suspended $\SiN$ nanobeam of thickness $t=350$~nm and width $w=760$~nm, with an etched array of elliptical holes as shown in Fig.~\ref{Fig_1}(a). In the outer mirror sections, the spacing $a$ between the holes is constant, while within the cavity section, it varies quadratically from the center outwards. The cavity section forms a defect in an otherwise perfect 1D photonic bandgap structure.
The lattice modulation is designed to support localized, high $Q$ ($Q_o>10^6$) transverse-electric (TE, with $\mathbf{E_z}=0$ on the $xy$ plane) optical modes in the 980~nm band, co-located with breathing-type mechanical resonances at $f_m\approx4$~GHz (respectively shown in Figs.~\ref{Fig_1}(b) and (c)). Figure~\ref{Fig_1}(d) shows a scanning electron microscope (SEM) image of a nanofabricated device, where it is clear that, in addition to the lattice spacing modulation, the aspect ratio of the holes is adjusted from the cavity region outwards.
As detailed below, the design procedure leading to this final geometry involved finite-element method determination of optical and mechanical resonances and the optomechanical coupling rate $g_0$ due to radiation-pressure. The latter was calculated from modal electromagnetic field and mechanical displacement profiles via a first-order perturbative expression described in the supplement~\cite{ref:SiN_nanobeam_note}, which only takes into account the moving-boundary contribution to the optomechanical coupling. Contributions from stress-induced changes of the refractive index (photoelastic effect) were not considered although they may be as significant as those in crystalline silicon nanostructures~\cite{ref:chan_optimized_OMC,ref:rakich_PRX}.

\begin{figure}[h]
\centerline{\includegraphics[width=9cm,trim=0 15 0 15]{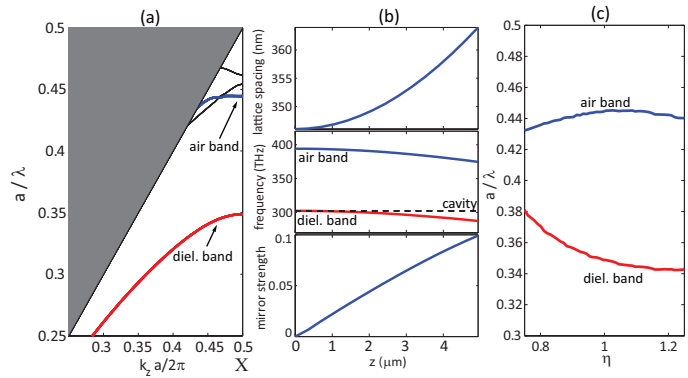}}
\caption{(a) TE-polarization photonic bands for a periodic array of holes. Dielectric and air band modes are of first order in the $x$ direction. Black curves: high order modes. (b) Hole spacing (top), $\mathbf{X}$-point dielectric and air band frequencies (middle) and effective mirror strength (bottom) as functions of distance $z$ along the cavity ($z=0$ is the cavity center) for a high $Q_o$ OMC cavity. (c) Evolution of the $\mathbf{X}$-point dielectric and air band frequencies as a function of the hole aspect ratio factor $\eta$.}
\label{Fig_2}
\end{figure}

We next detail our design procedure. TE Photonic bands for a 1D nanobeam photonic crystal are shown in Fig.~\ref{Fig_2}(a). Modes on the dielectric and air bands are of fist order in the transverse ($x$) direction (black lines correspond to higher transversal order modes). Parameters for the cavity section of the OMC (see Fig.~\ref{Fig_1}(a)) were obtained iteratively as follows. For fixed thickness $t=350$~nm and aspect ratio $w_x/w_z$, and an initial guess for the width $w$, a lattice constant $a$ was selected such that the edge of the dielectric band at the $\mathbf{X}$ point ($k_z=\pi/a$) fell close to the desired resonance wavelength. The width $w$ was varied, the $\mathbf{X}$-point dielectric band edge was recalculated, the lattice constant $a$ was adjusted, and the procedure was repeated until convergence of the dielectric band edge to the desired wavelength was reached. Once the lattice constant for the cavity section was determined, the effect of varying this parameter along the OMC (in the $z-$direction) was evaluated. As shown in Fig.~\ref{Fig_2}(b), increasing the lattice constant along $z$ while keeping all other dimensions fixed causes the dielectric band edge to red-shift, creating the condition for optical confinement by the photonic bandgap: the lattice constant is increased quadratically from the cavity center ($z=0$) towards the edges, so that the allowed frequency at $z=0$ (dashed line) falls within the bandgap of the outer regions (e.g., $z=5~\mum$). The cavity can also be regarded as a distributed reflector with a locally varying, distributed reflectivity as in the bottom panel in Fig.~\ref{Fig_2}(b), which is close to linear near $z=0$. Linear mirror strength profiles~\cite{ref:SiN_nanobeam_note} tend to produce optical modes with reduced spatial harmonics above the light line. This leads to
reduced power leakage into the air, and thus higher optical quality factors~\cite{ref:Quan1,ref:Srinivasan1}. Confined modes with optical quality factors in excess of $10^6$ were obtained simply with the lattice profile of Fig.~\ref{Fig_2}(c) and $w_z/w_x\approx0.5$. We note that, because $\SiN$ has a considerably smaller refractive index than Si (2.0 in contrast with 3.5), the achievable local reflectivities are correspondingly smaller, and thus in general the number of unit cells necessary for a high $Q_o$ cavity is larger, and so are the obtained mode volumes.

\begin{figure}[h]
\centerline{\includegraphics[width=9cm,trim=0 15 0 15]{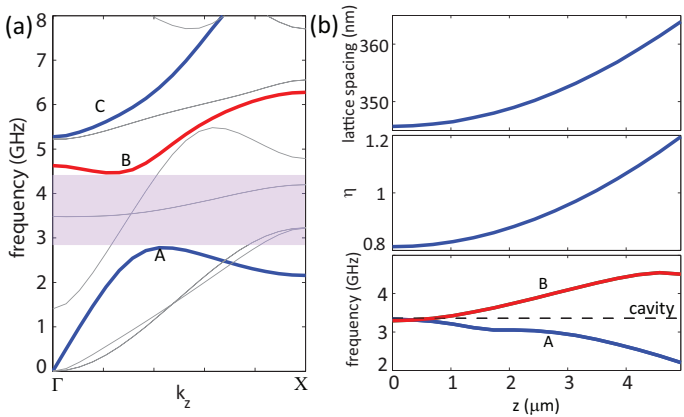}}
\caption{(a) Phononic bands for a periodic array of holes with $a=350$~nm. Modes on bands A, B and C are symmetric across planes $x$, $y$ and $z$. Modes on the gray bands have other types of spatial symmetries. (b) (top) Hole spacing, (middle) hole aspect ratio factor $\eta(z)$, and (bottom) A band maximum (blue) and B band minimum (red) frequencies as functions of distance $z$ along the cavity. Bottom panel, dashed line: fundamental breathing mode frequency.}
\label{Fig_3}
\end{figure}

The resulting 1D photonic crystal geometries support a phononic bandgap that is used to generate a localized mechanical mode. The band structure for the crystal shown in Fig.~\ref{Fig_2}(a) is plotted in Fig.~\ref{Fig_3}(a), for a lattice constant $a=350$~nm. Thick lines indicate spatially symmetric displacement across the $x$-$z$ and $y$- $z$-planes, gray lines to other symmetries. A phononic bandgap exists between the A and B bands (shaded region). Near the $\Gamma$-point, modes on the band B (red line) have a displacement pattern that gives rise to breathing resonances where the outer nanobeam walls expand or contract symmetrically. We found that the quadratic lattice constant modulation was insufficient to produce a well-confined mechanical mode, yielding only a relatively small shift of the mechanical bands. To overcome this, we allowed the minor and major radii of the elliptical holes to vary along the cavity as $(w_z(z),w_x(z))=(w_{z,0}/\eta(z),w_{x,0}\cdot\eta(z))$, with quadratic $\eta(z)$ (Fig.~\ref{Fig_3}(b) (middle)). This causes the strong modulation of the edges of bands A and B, see in Fig.~\ref{Fig_3}(b) (bottom), so allowed frequencies at $z=0$ (cavity center) fall within the phononic bandgap of the outer regions ($z>5~\mum$). The breathing mode frequency is indicated with a dashed line in Fig.~\ref{Fig_1}(c). The $\eta(z)$ profile of Fig.~\ref{Fig_3}(b) was obtained with a nonlinear optimization routine that sought to maximize the optomechanical coupling rate $g_0$, while keeping the optical quality factor above $10^6$. The hole aspect ratio modulation does cause the photonic band edges to shift as shown in Fig.~\ref{Fig_2}(c). High $Q_\text{o}$ optical modes can still be found, albeit at frequencies shifted from the original. The optimized OMC design yielded $g_0/2\pi=133.6$~kHz ($\lom=5.1~\mum$) for an optical mode at $\lambda =966$~nm.

\begin{figure*}[t]
\centerline{\includegraphics{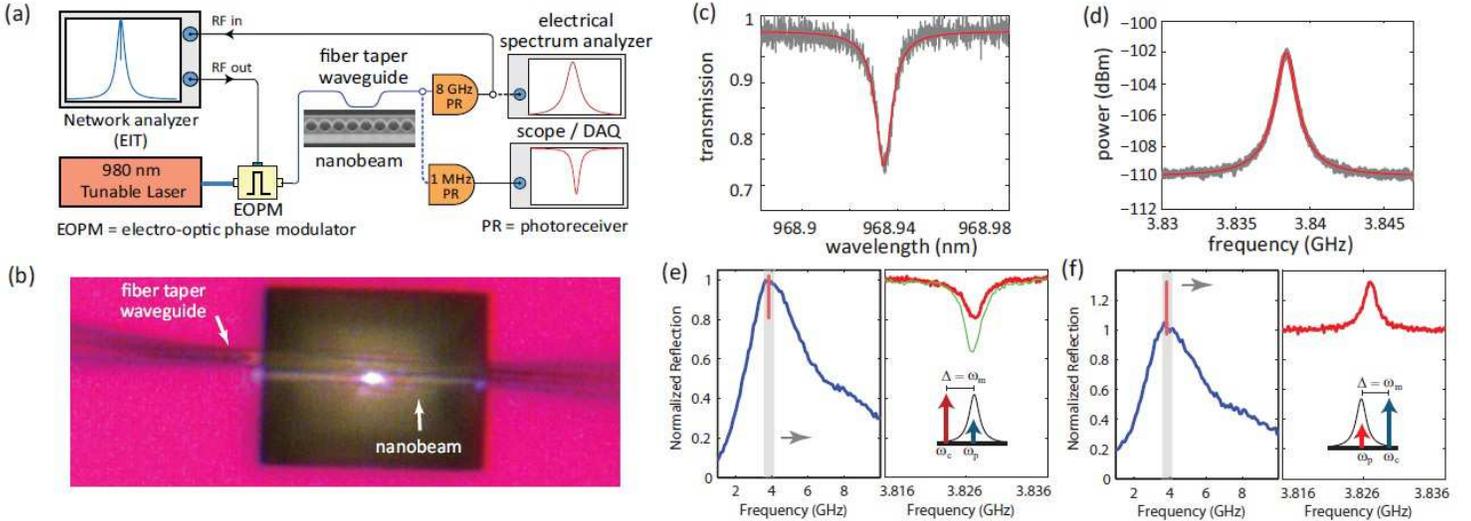}}
\caption{Experimental data. (a) Setup for optical and mechanical mode characterization. (b) Microscope image of fiber taper waveguide coupled to nanobeam cavity. DAQ = Data Acquisition. (c) Optical transmission spectrum showing resonance dip with $Q_o=116\times10^3\pm5\times10^3$ (d) Optical signal RF spectrum showing peak due to thermal motion of the fundamental breathing mechanical mode, with $Q_m=3152\pm30$. (e) Optical reflection spectrum for red-detuned pump $\approx\omega_m$ from the resonance center (see inset). Left panel: broad range spectrum; right panel: blow-up of reflection spectrum around the mechanical frequency, showing the EIT dip for CW pump. Green line: same, pulsed pump. (f) Same as (c), for blue-detuned pump. Left panel: reflection peak consistent with electromagnetically induced absorption. Errors in $Q_o$ and $Q_m$ are 95~\% fit intervals, corresponding to 2 standard deviations.}
\label{Fig_4}
\end{figure*}

Devices were fabricated on stoichiometric $\SiN$ (tensile stress of $\approx800$~MPa) with electron-beam lithography and reactive ion etching with a CHF$_3$:O$_2$:Ar mixture. Suspended nanobeams were obtained by etching the underlying Si substrate with KOH. Fabricated devices were initially characterized with optical transmission spectroscopy (Fig.~\ref{Fig_4}(a)). Light from a tunable external cavity diode laser was coupled to the devices using an optical fiber taper waveguide (FTW), as in Fig.~\ref{Fig_4}(a). A polarization controller before the FTW allowed coupling to TE cavity modes to be maximized. The laser wavelength was swept and the signal transmitted through the FTW was detected and recorded, revealing optical resonances in the 980 nm band with $10^4<Q<1.3\times10^5$ (Fig.~\ref{Fig_4}(c)). Mechanical modes coupled to a particular optical resonance were measured at low optical power by tuning the laser wavelength to the shoulder of the optical mode. Resonance fluctuations induced by thermal-noise-driven mechanical motion were converted to an intensity modulation of the transmitted optical signal. The transmitted optical signal was detected with an avalanche photodiode (APD) and the electrical signal was resolved in a spectrum analyzer. A typical spectrum showing a peak due to the fundamental breathing mechanical mode at $\approx 3.8$~GHz is shown in Fig.~\ref{Fig_4}(d). With a quality factor $Q_m\approx3000$ at atmosphere, the frequency-$Q_\text{m}$ product is $\approx12\times10^{12}$, twice that observed in vacuum and at 8K in ref.\cite{ref:Liu_yuxiang_wlc}. The high mechanical frequency ensures that the system is in the resolved-sideband regime ($\omega_m/\kappa\approx 1.6$).

The typical sideband resolution achieved was sufficient for the observation of optomechanical EIT at room temperature and atmospheric pressure.  This effect corresponds to the creation of a narrow transparency window in the optical transmission spectrum (and, correspondingly, a dip in reflection) by pumping the optical cavity at a red-detuned frequency from the cavity center. It is a macroscopic manifestation of the coherent interaction between cavity photons and phonons, as it stems from the destructive interference between an incident probe signal and anti-Stokes pump photons scattered by the mechanical resonance.  Optomechanical EIT has been observed in many systems, including silica microtoroids~\cite{ref:Weis_Kippenberg} and microspheres~\cite{ref:Dong_Wang_Science_dark_mode}, Si OMCs~\cite{ref:safavi-naeini4}, and $\SiN$ microdisks~\cite{ref:Liu_yuxiang_wlc}, though these previous demonstrations were typically done in vacuum (and in many cases, at cryogenic temperatures).

To measure this effect, an electro-optic phase modulator was used to produce sidebands on a continuous wave (CW) control beam. With the control field fixed at a frequency $\omega_c$, $\Delta$ away from the optical resonance at $\omega_o$, the modulation frequency was varied over the range 100~MHz to 20~GHz, which allowed the sidebands to scan over the entire optical resonance. As detailed in~\cite{ref:SiN_nanobeam_note}, detection of the modulated signal with the 8~GHz photoreceiver  and demodulation in a vector network analyzer produced the cavity reflectivity spectra in Fig.~\ref{Fig_4}(e). The left panel of Fig.~\ref{Fig_4}(e) shows the reflection spectrum over a broad frequency range, indicating an optical cavity FWHM of $\approx 4.4$~GHz. The red portion of the curve is the superimposed spectrum shown on the right panel, where the reflectivity dip due to EIT is clearly evident. Figure~\ref{Fig_4}(f) shows the same for a blue-detuned pump. Here, optomechanical electromagnetically induced absorption (EIA) manifests itself as a peak in reflection, corresponding to constructive interference between the probe and Stokes-scattered pump photons. Assuming $\Delta\approx\omega_m$ ($\omega_m$ is the mechanical frequency), a cooperativity $C=0.12$ was estimated via a fit to the EIT and EIA spectra~\cite{ref:SiN_nanobeam_note}. From the transmission spectrum and control laser detuning $\Delta$, we estimate an intracavity photon population of $N=(9.73\pm1.34)\times10^3$, and thence, an optomechanical coupling rate $g_0/2\pi=170$~kHz$\pm17$~kHz~\cite{ref:SiN_nanobeam_note}, higher than the theoretical $g_0/2\pi\approx130$~kHz (uncertainties are of one standard deviation). The discrepancy is possibly due to geometrical variations between the fabricated device and the simulated model. Another possibility is the existence of a constructive elasto-optic contribution~\cite{ref:chan_optimized_OMC}, which is not included in our model.

Linear absorption in the $\SiN$ leads to a red-shift of the optical resonance frequency (thermo-optic effect) which increases with cavity photon number and results in a bistable dynamic behavior~\cite{ref:BarclayOE05}. At sufficiently high powers, the shift reaches an unstable maximum, and the cavity snaps back to its original frequency. In recent work with $\SiN$ microdisks~\cite{ref:Liu_yuxiang_wlc}, these issues were mitigated by cryogenic cooling and actively locking the pump laser to the cavity. The achieved $C$ was nevertheless limited to 0.5 in spite of the larger intracavity photon number ($\approx 6\times10^4$) and higher optical and mechanical quality factors ($Q_\text{o}=5\times10^5$, $Q_\text{m} = 10^4$) than here. This was due to the optomechanical coupling rate, $g_{0}/2\pi\approx10$~kHz, ~$>10\times$ lower than in the $\SiN$ OMC. This comparison indicates that high $g_0$ is of practical importance, offsetting the effects of limited sideband resolution, mechanical quality factor, and thermo-optic dispersion. Here, achieving significantly higher $N$ and $C$ under CW pumping was limited by thermo-optic cavity dispersion and instability. To mitigate this, we performed EIT measurements with a pulsed pump~\cite{ref:SiN_nanobeam_note}, and were able to reach a maximum population of $N\approx3\times10^4$ ($3\times$ that for CW). A maximum $C=0.264\pm0.01$ was observed (green curve in Fig.~\ref{Fig_4}(e)). Evidence of nonlinear absorption was also observed.~\cite{ref:SiN_nanobeam_note}

In summary, we have designed a $\SiN$-based nanobeam optomechanical crystal supporting $Q_\text{o}>10^6$ optical resonances in the $980$~nm wavelength band and breathing mechanical modes in the 4~GHz range, with relatively high optomechanical coupling rates ($g_0/2\pi=130$~kHz). With high mechanical frequencies, nanobeam cavities with optical quality factors of $\approx10^5$ were operated in the sideband resolved regime, and were used to demonstrate electromagnetically induced transparency and absorption at room temperature and atmospheric pressure, with intracavity photon numbers on the order of $10^4$. The observed mechanical frequency-$Q_\text{m}$ product was $\approx12\times10^{12}$. Moving towards $C\gg1$ - as necessary for e.g., quantum level photon-phonon translation~\cite{ref:safavi-naeini3} - our results suggest that strategies for producing enhanced zero-point optomechanical coupling rates, such as slot-mode type OMCs~\cite{ref:Davanco_OMC} can be advantageous for $\SiN$ devices, which, unlike silicon devices, are able to support large intracavity photon numbers ($\approx10^4$) before starting to show signs of nonlinear absorption. Other wide bandgap materials, including AlN~\cite{ref:fan_hong_tang} and GaP~\cite{ref:Mitchell_arxiv_1309.6300v2} (which has refractive index $\approx$3) are promising candidates as well.

We thank Vladimir Aksyuk and Oskar Painter for helpful discussions.  This work was partially supported by the DARPA MESO program. S.A. acknowledges support under the Cooperative Research Agreement between the University of Maryland and NIST-CNS,T Award 70NANB10H193. Y. L. acknowledges support under the NIST-ARRA Measurement Science and Engineering Fellowship Program Grant No. 70NANB10H026 through the University of Maryland.




\onecolumngrid \bigskip
\appendix
\setcounter{figure}{0}
\setcounter{equation}{0}
\makeatletter
\renewcommand{\theequation}{S\@arabic\c@equation}
\begin{center} {{\bf \large SUPPORTING
INFORMATION}}\end{center}

\section*{Computational Models}

Photonic crystal band structures were obtained with the plane-wave expansion method, using supercells in the $x$ (lateral) and $y$ (vertical) directions.
Optical resonant modes were calculated by solving the vector eigenvalue wave equation for the electric field with the finite element method. Vector edge elements were used, and a radiation boundary condition at the outer edges to simulate open space. Due to the radiative boundary conditions, the frequency $\omega$ was complex, which allowed the calculation of the optical quality factor as $Q=\text{Re}\{\omega\}/(2\cdot\text{Im}\{\omega\})$.

Mechanical resonances were obtained by solving the elastic wave equation for mechanical displacement, assuming $\SiN$ to be an isotropic material~\cite{ref:eichenfield1_} and the two ends of the nanobeam to be perfectly clamped.

\subsection*{1D photonic crystal mirror strength}
The effective mirror strength for the 1D photonic crystal lattice, mentioned in the main text, corresponds to the imaginary part of the Bloch
wavenumber for bandgap frequencies at the Brillouin zone edge,
$k=\pi/a(1+i\gamma)$. From 1D first order perturbation theory,
\begin{equation}
\gamma = \sqrt{\left(\frac{\omega_2-\omega_1}{\omega_1+\omega_2}\right)^2
-\left(\frac{\omega-\omega_0}{\omega_0}\right)^2},
\end{equation}
where $\omega_{1,2}$ are the dielectric and air band edges at the
Brillouin zone boundary, and $\omega_0$ is the midgap frequency.
Linear mirror strength profiles tend to produce optical modes with
reduced spatial harmonics above the light line. This leads to
reduced power leakage into the air, and thus higher quality
factors~\cite{ref:Quan1,ref:Srinivasan1_}.

\subsection*{Optomechanical coupling}
The shift in the frequency $\omega_{o}$ of a
particular optical resonance due to displacement of the
nanostructure boundaries produced by a mechanical resonance at
frequency $\omega_m$ is quantified by the optomechanical coupling $g_{om} =
\partial \omega_o/\partial x=\omega_o/L_{\text{OM}}$; here, $x$ is the cavity
boundary displacement and $L_{\text{OM}}$ is an effective optomechanical interaction
length~\cite{ref:eichenfield2_}. The effective length $L_{\text{OM}}$ can be
estimated via the perturbative expression
\begin{equation}
\label{eq:Lom} L_{\text{OM}} = \frac{2\int {dV \epsilon\left|\mathbf{E}
\right|^2}}{\int{dA\left(\mathbf{Q}\cdot \mathbf{n}
\right)\left( \Delta \epsilon\left| \mathbf{E}_\parallel
\right|^2-\Delta(\epsilon^{-1}) \left| \mathbf{D}_\perp \right|
^2\right)} }.
\end{equation}

Here, $\mathbf{E}$ and $\mathbf{D}$ are the modal electric and
electric displacement fields, respectively, $\Delta \epsilon =
\epsilon_{diel.}-\epsilon_{air}$, $\Delta(\epsilon^{-1}) =
\epsilon_{diel.}^{-1}-\epsilon_{air}^{-1}$, and $\epsilon_{diel.}$ and $\epsilon_{air}$ are the permittivities of the nanobeam material and air, respectively. The
mass displacement due to the mechanical resonance is given by
$\mathbf{Q}$, and the normal surface displacement at the structure
boundaries is $\mathbf{Q}\cdot \mathbf{n}$, where
$\mathbf{n}$ is the surface normal. The integral in the denominator
is performed over the entire surface of the nanostructure.

The optomechanical coupling $g_{om}$ was converted into the
coupling rate $g_0$ between photons and phonons by
$g_0=x_{zpf}\cdot g_{om}$, where $x_{zpf}=\sqrt{\hbar/2m\omega_m}$ is
the zero point fluctuation amplitude for mechanical
displacement and $m$ is the motional mass
of the mechanical resonance at frequency $\omega_m$. The motional
mass was obtained from the displacement $\mathbf{Q}$ and the
nanobeam material density $\rho$ by
$m=\rho\int{dV\left(\frac{|\mathbf{Q}|}{\text{max}(|\mathbf{Q}|)}\right)^2}$~\cite{ref:safavi-naeini3_}.

\section*{Device fabrication}

On a silicon substrate, a 350~nm thick stoichiometric layer of Si$_3$N$_4$ was grown by low pressure chemical vapor deposition (LPCVD). The film had a process-induced internal tensile stress of $\approx800$~MPa, measured by the wafer bowing method. A 500~nm thick positive-tone electron beam (E-beam) resist was spin-coated on the Si$_3$N$_4$ film, followed by E-beam lithography and development in hexyl acetate at 8~$^{\circ}$C. The patterns were transferred into the Si$_3$N$_4$ layer by an O$_2$/CHF$_3$/Ar reactive ion etch (RIE). After resist removal using a stabilized H$_{2}$SO$_{4}$/H$_{2}$O$_{2}$ solution, the sample was immersed in a $75~^\circ$C, 45~$\%$ KOH bath for approximately 15 min, to etch away the Si substrate, thereby releasing the $\SiN$ nanobeam. The sample was then placed in a 1:4 H$_2$O:HCl solution for 5 min. to remove precipitates from the KOH etching, then rinsed and N$_2$ blow-dried.

\section*{Experimental setup}

\subsection*{Optical and mechanical mode spectra}
The experimental setup for device characterization and EIT measurements is shown in Fig.4(a) in the main text. Light from a 980~nm band tunable diode laser is evanescently coupled into and out of the devices using an optical fiber taper waveguide (FTW), as shown in Fig.4(b). Optical modes are measured by sweeping the laser wavelength, detecting the transmitted light with a 1~MHz bandwidth photoreceiver and recording the detected signal. A fiber polarization controller (FPC - not shown in Fig.4(a)) is used just before the FTW to ensure that only the TE polarization cavity mode is accessed. This is done by minimizing the transmission at the resonance center with the FPC.
Mechanical modes are measured with the laser wavelength tuned to the shoulder of an optical mode, by detecting the transmitted light with a 8~GHz bandwidth photoreceiver. Fluctuations in the transmitted power due to thermal nanobeam mechanical motion are spectrally resolved on a real-time electronic spectrum analyzer (RSA).
\subsection*{Electromagnetically Induced Transparency (EIT)}
In the EIT measurement setup, shown in Fig.4(a) of the main text, light from the 980~nm tunable laser corresponds to the control field, and a probe signal is derived from it by modulation with an electro-optic phase modulator (EOPM). This produces out-of-phase blue and red sidebands at frequencies $\pm\Delta_\text{pc}$ away from the control beam frequency $\omega_c$. As shown in Fig.4(a) of the main text, the EOPM is driven by port 1 of a vector network analyzer (VNA), so that the probe-control beam detuning $\Delta_{\text{pc}}$ can be swept. For small phase modulation index $\beta$, the optical signal fed into the FTW can be represented by
\begin{equation}
E_\text{in} = e^{i\omega_ct}+\frac{\beta}{2}\left[ e^{i(\omega_c+\Delta_\text{pc})t} - e^{i(\omega_c-\Delta_\text{pc})t} \right].
\end{equation}
After the cavity, which has a transmission transfer function $t(\omega)=|t(\omega)|e^{i\phi(\omega)}$, this becomes
\begin{equation}
E_\text{out} = e^{i\omega_ct} \left\{ t(\omega_c) + \frac{\beta}{2} \left[ t(\omega_c+\Delta_\text{pc}) e^{i\Delta_\text{pc}t} - t(\omega_c-\Delta_\text{pc}) e^{-i\Delta_\text{pc}t} \right] \right\},
\end{equation}
The transmitted signal is then photodetected, yielding a photocurrent proportional to $|E_\text{out}|^2$. For sufficient sideband resolution, we may assume that $t(\omega_c-\Delta_\text{pc})\approx1$, i.e., red sideband does not see the cavity. Disregarding terms proportional to $\beta^2$ and using the relation $t(\omega)=1+r(\omega)$, where $r(\omega)$ is the cavity reflection coefficient~\cite{ref:safavi-naeini4_}, the photocurrent component at the modulation frequency $\Delta_\text{pc}$ is proportional to
\begin{equation}
\left| t(\omega_c) \right|\left\{ \left| r(\omega_c+\Delta_\text{pc}) \right| \cos\left[ \Delta_\text{pc}t-\phi(\omega_c)+\phi(\omega_c+\Delta_\text{pc}) \right] + 2\sin(\Delta_\text{pc}t)\sin[\phi(\omega_c)] \right\}.
\end{equation}
For $\phi(\omega_c)\to 0$, this becomes
\begin{equation}
\left| t(\omega_c) \right| \left| r(\omega_c+\Delta_\text{pc}) \right| \cos\left[ \Delta_\text{pc}t+\phi(\omega_c+\Delta_\text{pc}) \right].
\end{equation}
Under these conditions, the network analyzer $S_{21}(\Delta_\text{pc})$ parameter has amplitude proportional to $\left|r(\omega_c+\Delta_\text{pc})\right| $ and phase $\phi(\omega_c+\Delta_\text{pc})$.

\section*{Optomechanical coupling rate estimate}
\begin{figure}[h]
\centerline{\includegraphics[width=12cm,trim=0 0 0 0]{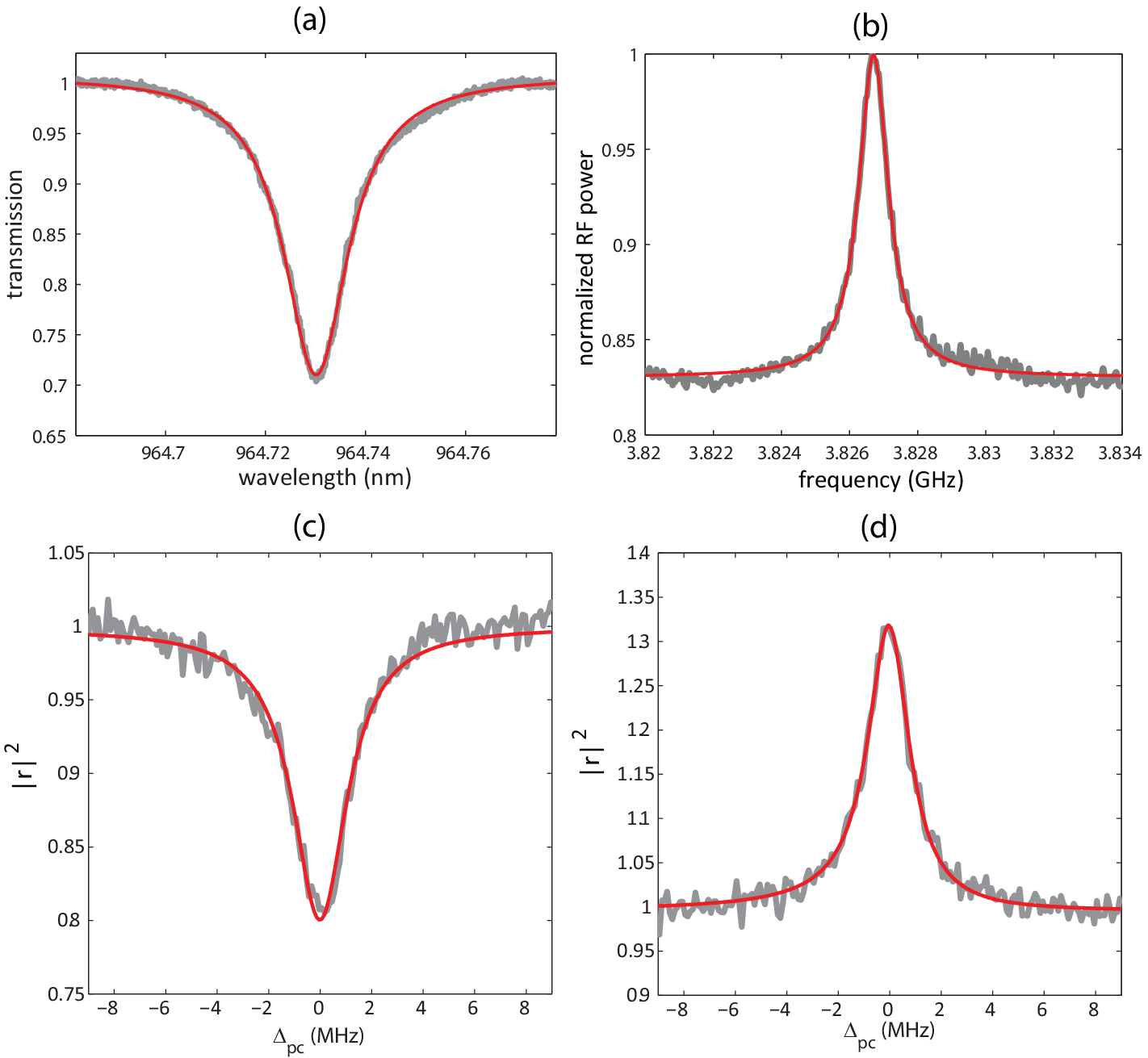}}
\caption{(a) Optical transmission and (b) RF mechanical spectra for the $\SiN$ nanobeam. (c) Electromagnetically induced transparency (EIT) and (d) absorption (EIA) traces as a function of the control-probe detuning $\Delta_\text{pc}$. Grey lines: data. Red lines: fits.}
\label{Fig_S1}
\end{figure}

The optomechanical cavity reflectivity coefficient $|r(\omega)|$ can be cast as a function of the control-probe detuning $\Delta_\text{pc}$ as

\begin{equation}
r(\Delta_{\text{pc}}) = -\frac{1}{1+\frac{2i(\Delta_{\text{oc}}-\Delta_{\text{pc}})}{\kappa}+\frac{C}{\frac{2i(\omega_{\text{m}}-\Delta_{\text{pc}})}{\gamma_{\text{m}}}+1}},
\label{eq:Seq1}
\end{equation}
where $\kappa$ is the optical cavity decay rate, $\Delta_\text{oc}$ cavity-control detuning, $\omega_m$ the mechanical frequency, $\gamma_m$ the intrinsic mechanical damping and $C$ the cooperativity.
This expression is used to fit the EIT reflectivity data shown in Fig.4(e) in the main text, with $C$, $\Delta_\text{oc}$ and $\gamma_m$  as fit parameters. The parameter $\kappa$ is estimated by taking the full width at half maximum of the broad range reflectivity spectrum in Fig.4(e), $\kappa/2\pi=4.03$~GHz$\pm0.06$~GHz. At the same time, the broad range reflectivity peak  in Fig.4(e) has a FWHM of $\approx4.4$~GHz, so we use $\kappa/2\pi=4.2$~GHz$\pm0.2$~GHz. From the maximum of the mechanical mode spectrum (Fig.~\ref{Fig_S1}(b)), we obtain $\omega_m=3.83$~GHz. From the fit to the EIT curve (Fig.~\ref{Fig_S1}(c)), we obtain $C=0.118~\pm 0.004$, $\Delta_\text{oc}/2\pi=3.88$~GHz$~\pm 0.04$~GHz and $\gamma_m/2\pi=2.3$~MHz $~\pm$ 0.1~MHz. The uncertainties correspond to 95~\% fit confidence intervals and are due to experimental noise in the collected data. Applying the same procedure to the EIA curve (Fig.~\ref{Fig_S1}(d)), we obtain $C=0.131~\pm 0.002$, $\Delta_\text{oc}/2\pi=3.98$~GHz$~\pm 0.04$~GHz and $\gamma_m/2\pi=2.3$~MHz~$\pm$ 0.1~MHz.

To estimate the zero-point optomechanical coupling rate $g_0$, we recall that $C=4Ng_0/\kappa\gamma_m$, where $N$ is the intracavity photon number:
\begin{equation}
N = \frac{1}{\hbar\omega_{o}}\eta\Delta TQ_{i}\left(\frac{P_{\text{in}}}{\omega_o}\right)\frac{1}{1+(\frac{\Delta_{\text{oc}}}{\kappa/2})^2}.
\label{eq:Seq2}
\end{equation}

Here, $\hbar=h/2\pi$ ($h$ is Planck's constant), $\eta$ is the FTW coupling efficiency, $\Delta T$ is the depth of the optical resonance in the transmission spectrum ($\Delta T$=1 at critical coupling), $Q_i$ the intrinsic optical $Q$, and $P_{\text {in}}=1$~mW is the optical power at the FTW input. The intrinsic $Q$ was obtained as $Q_i= 2Q /(1+\sqrt{1-\Delta T})$. From the optical transmission spectrum (Fig.~\ref{Fig_S1}(a)), we obtained $\Delta T = 0.30 \pm 0.01$. Here, the uncertainty corresponds to twice the standard deviation of the 100 first transmission values at the short wavelength limit in Fig.~\ref{Fig_S1}(a). The mean transmission value for the same 100 samples was used to normalize the measured spectrum.

The FTW coupling efficiency $\eta$ models intrinsic losses that arise in the tapered fiber fabrication, as well as external factors that may reduce the probe transmission - e.g., dust particles on the FTW surface that may scatter the guided light before it reaches the cavity. A measurement of transmitted power through the FTW away from the resonance revealed a power transmission $T_\text{FTW}\approx35~\%$. Optical transmission spectra measured in the forwards and backwards directions (i.e., with the tunable laser coupled to either ports 1 or 2 of the FTW), however, revealed different bistable behaviors at high laser powers. Because bistable behavior is a function of the power dropped into the cavity, that observation suggested that the power coupled from the FTW into the cavity depended on whether cavity excitation was from port 1 or 2. We thus postulate that the total FTW transmission is $T_\text{FTW}=T_1\times T_2$, where $T_{1,2}$ is the transmission from port 1 or 2 of the FTW up to the cavity. If coupling into the the FTW is asymmetric from ports 1 or 2, we can write $T_1/T_2 = r$, so that $T_1=\sqrt{r T_\text{FTW}}$. To determine $\eta$ in eq.~(\ref{eq:Seq2}) for the direction in which the EIT and EIA curves were obtained (i.e., from port 1 to 2: $\eta=T_1$), a bistable transmission spectrum was measured in the forward direction, with a known laser power $P_\text{f}$. The excitation was then switched to port 2, and transmission spectra were taken at varying powers $P_\text{b}$ until the same bistable behavior as observed in the forward direction was obtained. This happened at $P_\text{f}/P_\text{b}\approx1.5\equiv r$, so that $\eta\approx\sqrt{0.35\cdot1.5}=0.72$. Using these parameters in eq.~\ref{eq:Seq2}, we obtain $N=9730\pm1340$. The uncertainty in $N$ was obtained by propagating the uncertainties for the different parameters.

Substituting all values in eq.~\ref{eq:Seq2}, we are able to estimate the optomechanical coupling rate to be $g_0/2\pi=170$~kHz$\pm17$~kHz. Applying the same procedure to the EIA curve, we estimate $N=9268\pm1235$ and $g_0/2\pi=183$~kHz$~\pm 17$~kHz.

\section*{Pulsed pump measurements}
\begin{figure}[h!]
\centerline{\includegraphics[width=17cm,trim=0 0 0 0]{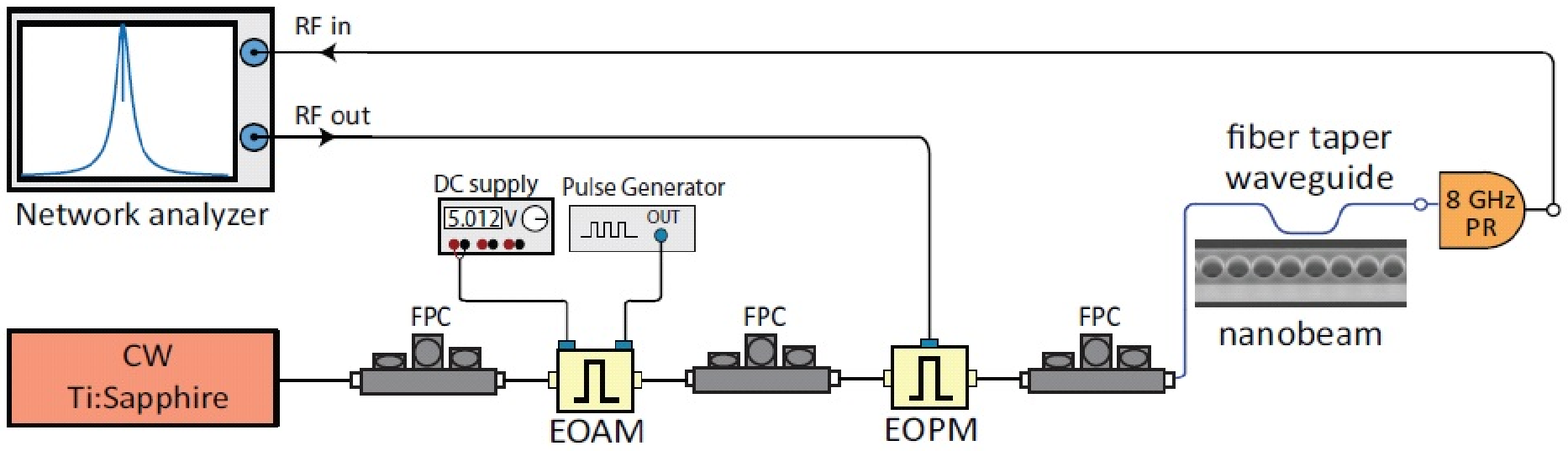}}
\caption{Pulsed pump EIT measurement setup. EOAM:electro-optical amplitude modulator. EOPM:electro-optical phase modulator; PR:photoreceiver; FPC:fiber polarization controller.}
\label{Fig_S2}
\end{figure}

\begin{figure}[h!]
\centerline{\includegraphics[width=18cm,trim=0 0 0 0]{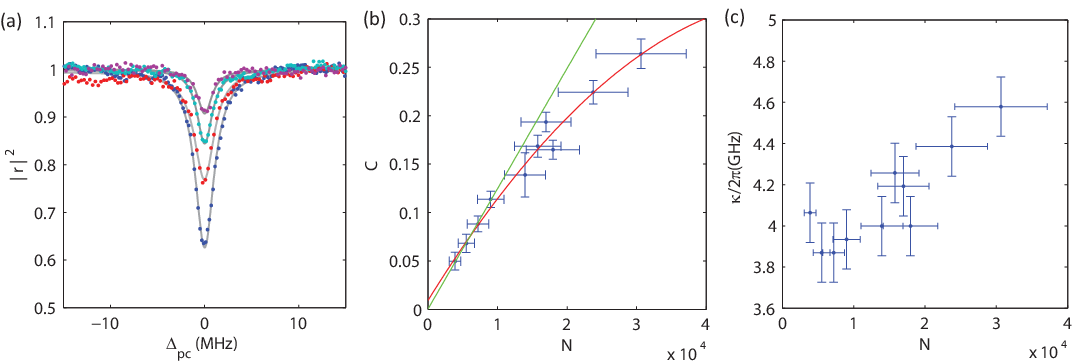}}
\caption{(a) EIT signal for varying intracavity photon number N. Circles: experimental data. Grey lines: fit. (b) Cooperativity (C) as a function of intracavity photon number $N$. Green line: linear fit to the first four points ($N<1\times10^4$). Red line: quadratic fit to the entire dataset. (c) Optical linewidth ($\kappa/2\pi$) as a function of $N$.}
\label{Fig_S3}
\end{figure}

In order to achieve higher intracavity photon population numbers, we repeated the EIT measurements described above with a pump signal consisting of a periodic sequence of short, rectangular optical pulses. The duty cycle and pulse width were chosen to minimize cavity heating by linear absorption of the pump, and thereby the thermo-optic cavity dispersion. The experimental setup is illustrated in Fig.~\ref{Fig_S2}.
Light from a CW Ti:Sapph laser was initially passed through an electro-optic amplitude modulator (EOAM) to produce a 100~kHz periodic train of 1~$\mu$s pulses with an extinction $R_\text{dB}$ such that $11.8$~dB $\leq R_\text{dB} \leq16.3$~dB. The uncertainty here is due to noise in the high and low power intensities detected with the 1~MHz photoreceiver and observed in an oscilloscope trace. The amplitude-modulated signal was then phase modulated with an electro-optic phase modulator (EOPM) driven by port 1 of the Vector Network Analyzer (VNA), and was launched into the FTW, which was evanescently coupled to the nanobeam OMC. Output light from the FTW was detected with a 8~GHz photoreceiver, the RF output of which was connected to port 2 of the VNA. The VNA displayed the $S_{21}$ parameter for the system, which, as shown above, is proportional to the cavity reflectivity signal. The VNA IF filter bandwidth was set to 10~kHz, which removed the high frequency signal components due to the amplitude modulation. It is worth noting that pulses shorter than $1~\mu$s had bandwidth comparable to the EIT linewidth, which led to a low-pass filter broadening of the EIT dip.
Figure~\ref{Fig_S3}(a) shows EIT dips obtained for various input pump powers, and pump-cavity detuning $\Delta_\text{pc}\approx\omega_m$. The corresponding intracavity photon numbers are indicated in the figure. Fitting each EIT curve as detailed above allowed us to produce the curve in Fig.~\ref{Fig_S3}(b) for the cooperativity as a function of intracavity photon number. The highest intracavity photon of $\approx3\times10^4$ is achieved is limited by thermo-optic dispersion, which causes the cavity to become unstable for the desired detuning. We observe that the cooperativity increases linearly with $N$ from $\approx 0.5\times10^4$ to $\approx1\times10^4$, however the full data set is better fitted with a quadratic polynomial. At the same time, the FHWM of the broad cavity reflectivity spectra (Fig~.\ref{Fig_S3}(c))) remains roughly at the same level (within measurement error) at the lower powers, going above it for all but the two highest powers. These results suggest that nonlinear absorption may be taking place for intracavity populations above $1\times10^4$.

\bibliographystyle{apsrev}

\end{document}